\documentclass[journal=nalefd,manuscript=letter,print]{achemso}
\usepackage{epsf}
\usepackage{graphicx}
\usepackage{dcolumn}
\usepackage{bm}
\usepackage{amsmath}
\usepackage{amsfonts}
\usepackage{amssymb}
\usepackage{ulem}
\usepackage{color}

\title{Anatomy and giant enhancement of the perpendicular magnetic anisotropy of cobalt-graphene heterostructures}

\author{Hongxin Yang}
\affiliation{Univ. Grenoble Alpes, INAC-SPINTEC, F-38000 Grenoble, France; CNRS, SPINTEC, F-38000 Grenoble, France; and CEA, INAC-SPINTEC, F-38000 Grenoble, France}

\author{Anh Duc Vu}
\affiliation{CNRS, Inst. NEEL, F-38000 Grenoble, France; Univ. Grenoble Alpes, Inst. NEEL, F-38000 Grenoble, France}

\author{Ali~Hallal}
\affiliation{Univ. Grenoble Alpes, INAC-SPINTEC, F-38000 Grenoble, France; CNRS, SPINTEC, F-38000 Grenoble, France; and CEA, INAC-SPINTEC, F-38000 Grenoble, France}

\author{Nicolas Rougemaille}
\affiliation{CNRS, Inst. NEEL, F-38000 Grenoble, France; Univ. Grenoble Alpes, Inst. NEEL, F-38000 Grenoble, France}

\author{Johann Coraux}
\affiliation{CNRS, Inst. NEEL, F-38000 Grenoble, France; Univ. Grenoble Alpes, Inst. NEEL, F-38000 Grenoble, France}

\author{Gong Chen}
\affiliation{National Center for Electron Microscopy, Lawrence Berkeley National Laboratory, Berkeley, California 94720, USA}

\author{Andreas K. Schmid}
\affiliation{NCEM, Molecular Foundry, Lawrence Berkeley National Laboratory, Berkeley, California 94720, USA}

\author{Mairbek Chshiev}
\affiliation{Univ. Grenoble Alpes, INAC-SPINTEC, F-38000 Grenoble, France; CNRS, SPINTEC, F-38000 Grenoble, France; and CEA, INAC-SPINTEC, F-38000 Grenoble, France}
\email{mair.chshiev@cea.fr}

\begin{document}



\begin{abstract}
We report strongly enhanced perpendicular magnetic anisotropy (PMA) of Co films by graphene coating from both first-principles and experiments.
Our calculations show that graphene can dramatically boost the surface anisotropy of Co films up to twice the value of its pristine counterpart and can extend the out-of-plane effective anisotropy up to unprecedented thickness of 25~\AA.
These findings are supported by our experiments on graphene coating on Co films grown on Ir substrate.
Furthermore, we report layer-resolved and orbital-hybridization-resolved anisotropy analysis which help understanding the physical mechanisms of PMA and more practically can help design structures with giant PMA. As an example, we propose super-exchange stabilized Co-graphene heterostructures with a robust out-of-plane constant effective PMA and linearly increasing interfacial anisotropy as a function of film thickness.
These findings point towards possibilities to engineer graphene/ferromagnetic metal heterostructures with giant magnetic anisotropy more than 20 times larger compared to conventional multilayers,
which constitutes a hallmark for future graphene and traditional spintronic technologies.
\end{abstract}

\maketitle

Ferromagnetic (FM) electrodes possessing magnetic easy axis perpendicular to the interface is of major scientific interest due to their potential application for realizing next generation of spintronic devices including high density non-volatile memory and logic chips with high thermal stability~\cite{roadmap}. The traditional approach for perpendicular magnetic anisotropy (PMA) engineering is to use FM/oxide interfaces~\cite{Monso,Ohno} or multilayer structures comprising two FM or FM/nonmagnetic metal interfaces~\cite{mangin,beach}. Here, we propose a Co-graphene-based system to realize electrodes with giant PMA which might be of high importance for both traditional~\cite{wolf} and graphene~\cite{graphenespintronics} spintronics. 
The choice of graphene is governed by its passivating properties inherited from its high chemical inertness and impermeability~\cite{CoGraphene2010, 2012JPCL,2012APL}, its demonstrated outstanding physical properties\cite{GrapheneRMPgeim,risegraphene}, its very long spin diffusion length\cite{Tombros2007,Popinciuc2009, FertAPL,Han2011, Yang2011,Maassen2012,Dlubak2012} and the perfect tunneling spin-filtering effects it yields when sandwiched between two ferromagnetic electrodes~\cite{Kelly,GrTMR, GrNiTMR, GrMTJ,CoGrAl2O3}.
Therefore, if large PMA can be realized, graphene-Co structure can be very promising for the STT-MRAM applications \cite{Monso,Ohno,CoGraphene2010} in the sense that the device size can be strongly reduced since graphene is only single atomic layer thick. 

However, the fabrication of heterostructures made of a large area ($mm^2$), high-quality graphene layer in contact with a ferromagnetic metal thin film is challenging. The reason for that is twofold. First, most of transition metals usually form disordered assemblies of three-dimensional objects when deposited onto graphene or graphite surfaces \cite{Binns1999, Diaye2009}. Second, graphene growth on metals generally requires high-temperature processes that can induce dewetting of the thin film or intermixing between the thin film metal and its substrate. Several routes have been proposed to prepare atomically-flat hybrid interfaces that extend over large areas. For example, pulsed laser deposition of metals on graphene has been shown to be powerful to synthetize ferromagnetic metal/graphene interfaces \cite{CoGraphene2010}, while intercalation mechanism \cite{Tontegode1991} appears as a promising approach to fabricate graphene/ferromagnetic metal interfaces~\cite{Sicot2012, 2012APL, Decker2013, Vlaic2014, Vita2014, Pacile2014, Shick2014, Decker2014, Bazarnik2015}.

A broad family of graphene/metal intercalated systems has been explored to date, including those with a ferromagnetic metal. They exhibit properties that are not found in pristine graphene, such as a proximity-induced magnetism \cite{Weser2010, Weser2011} or a sizable spin-orbit interaction \cite{Varykhalov2008} giving rise in some cases to a topologically nontrivial electronic state~\cite{Calleja2014}. In these systems, graphene also plays an active role, modifying the properties of the intercalated metal compared to the case without graphene. When the metal is ferromagnetic, graphene, like carbon-based molecules~\cite{Repain2015}, was shown to affect the surface magnetic anisotropy of magnetic films~\cite{2012APL, 2012JPCL}. Besides, as an ultimately-thin capping materials, graphene passivates the surfaces of metals, rendering them almost insensitive to air oxidation. Accordingly 'graphene-passivated ferromagnetic electrodes' are considered as promising building-blocks in future spintronics devices~\cite{Dlubak2012}.

In this Letter, we revisited magnetocrystalline anisotropy of graphene-Co structures from both first-principles calculations and experiments. We demonstrate that graphene coating on Co films can dramatically enhance the PMA up to twice that of pristine Co films value depending on Co thickness. Moreover, graphene can increase the film effective PMA and stabilize an out-of-plane magnetization easy axis ferromagnetic layer thickness up to 25~\AA, which is much larger than that of the intensively studied Fe/MgO structure~\cite{Ohno,pmaMgO}.
In addition, our layer-resolved analysis reveals that the interfacial three Co layers play a decisive role in system's anisotropy and can be dramatically affected by the proximity of graphene. 
Furthermore, we provide the orbital-hybridization-resolved PMA analysis, from which we unveil the largest enhancement of PMA origins from a reversal of anisotropy of hybridization between $d_{z^2}$ and $d_{yz}$ orbitals due to graphene coating on Co. 
Finally, based on the anatomy of Co-graphene PMA, we propose Co-graphene heterostructures stabilized by super-exchange interaction, which are demonstrated by our first-principles calcualtions to possess a linearly increasing surface anisotropy and constant effective anisotropy as a function of film thickness.
\begin{figure}
 \includegraphics[width=\textwidth]{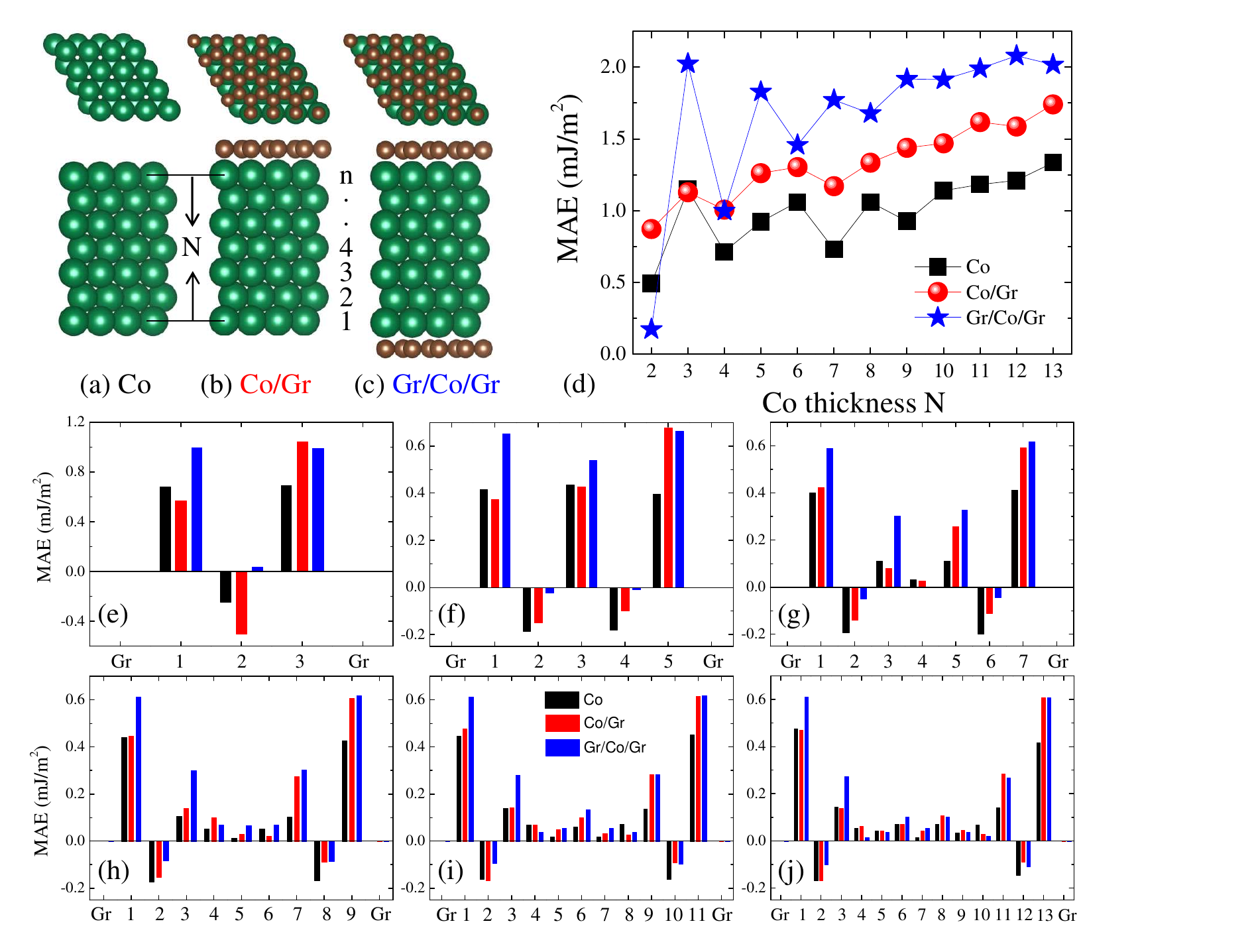}
\caption{Top and side view of bare Co slab (a), Co on Graphene (b) and Gr/Co/Gr (c), respectively. (d) Magnetocrystalline anisotropy energy as a function of  Co thickness N (monolayers). (e) to (j) are layer resolved local magnetocrystalline anisotropy energies for the cases with odd number of Co monolayers.}\label{fig1}
\end{figure}

Our first-principles calculations were carried out by using the Vienna $ab$~$initio$ simulation package (VASP)~\cite{vaspPRB93,vaspPRB96,vaspCMS}, where the electron-core interactions were described by the projector augmented wave method for the pseudopotentials~\cite{pawpotential}, and the exchange correlation energy was calculated within the generalized gradient approximation of the Perdew-Burke-Ernzerhof (PBE) form~\cite{gga,pbe}. 
The cutoff energies for the plane wave basis set used to expand the Kohn-Sham orbitals were 520 eV for all the calculations. 
$\Gamma$ centered 21 $\times$ 21 $\times$ 1 K-point mesh within Monkhorst-Pack scheme was used for the Brillouin zone integration. 
The PMA calculations were performed in three steps. First, structural relaxation was done until the forces are smaller than 0.001 eV/\AA ~for determining the ground state of each geometries for a bare Co film [Fig.~\ref{fig1}(a)], a Co film with one surface coated by graphene, Co/Gr [Fig.~\ref{fig1}(b)], and a Co film with both surfaces coated by graphene, Gr/Co/Gr [Fig.~\ref{fig1}(c)], respectively.
Next, the Kohn-Sham equations were solved with no spin-orbit coupling (SOC) taken into account to find out the charge distribution of the system's ground state.
Finally, the SOC was included and the non-self-consistent total energy of the system was determined when the orientation of the magnetic moments were set in-plane and out-of-plane, respectively. 

By comparing the total energy difference between in-plane and out-of-plane magnetic orientations, we obtained the magnetocrystalline anisotropy energy (MAE)~\cite{pmaMgO}. The results are shown in [Fig.~\ref{fig1}(d)]. The black squares are for a bare Co films with varying its thickness $N$ from 2 to 13 monolayers (ML).
One can see that the PMA of bare Co films is oscillating when the film thickness is less than 10 MLs due to the confined quantum well states formed between symmetric top and bottom surfaces.\cite{LiPRB,LiIEEE} 
When Co is thickner than 10 MLs, the oscillations almost vanish but the surface anisotropy still slightly increases. The origin of this small increase is due to the out-of-plane anisotropy contributions from Co bulk layers, which can be seen from layer-resolved MAE in Fig.~\ref{fig1}(g)-(j).
For the Co film with only one surface covered by graphene shown in Fig.~\ref{fig1}(b), one carbon atom of graphene unit cell is located on top of a Co atom with distance about 2.1~\AA~ while another carbon atom is located on the hollow site, which is consistent with previous studies~\cite{Kelly}. 
Thanks to this single atomic layer of graphene coating, the PMA [red full circles in Fig.~\ref{fig1}(d)] is strongly enhanced for all thicknesses except for 3 ML of Co~\cite{CoGraphene2010}.
From layer-resolved MAE analysis [Fig.~\ref{fig1}(e)-(j)], we see that the dominating enhancement of PMA originates from three interfacial Co layers, i.e. the enhancement of PMA from the 1$^{st}$ and 3$^{rd}$ interfacial Co layers, as well as the reduction of in-plane anisotropy of the 2$^{nd}$ interfacial Co layer. Further from the interface, the contribution to total anisotropy due to graphene coating becomes much weaker.
\begin{figure}
 \includegraphics[width=\textwidth]{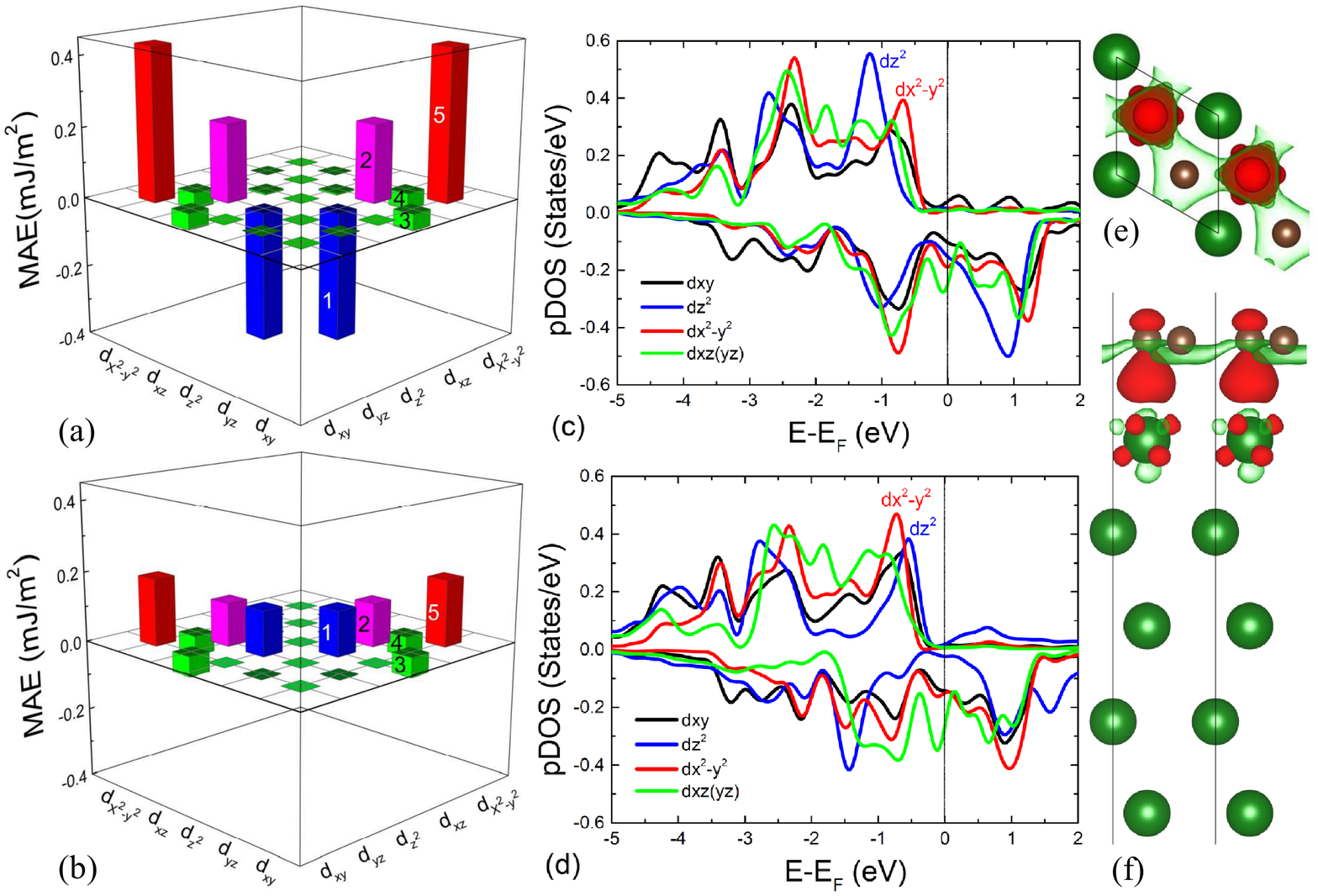}
 \caption{Magnetocrystalline anisotropy energy contributions from different orbital hybridizations at interfacial Co atom for a bare Co(5 ML) (a) and Co(5 ML)/graphene (b). Due to graphene coating, the hybridization between $d_{z^2}$ and $d_{yz(xz)}$ (blue bars labeled as 1) contributes to PMA (b) instead of in-plane anisotropy of pure Co case (a). The origin of this large anisotropy change can be attributed to the graphene coating caused redistribution of electronic density(electronic structure) on $3d$ orbitals of surface Co as shown by the projected density of states in (c) and (d) for pure Co and Co/Gr, respectively. The bonding between Co and graphene can also be seen from the charge difference, calculated by $\Delta \rho$=$\rho$(Co/Gr)-$\rho$(Co)-$\rho$(Gr), as shown in the top- (e) and side-view (f) of the contour charge difference when $\Delta \rho=\pm 4\times10^{-3}e/$\AA$^3$. The solid red clouds represent the charge accumulation, while the transparent green clouds represent the charge depletion.} \label{fig2}
\end{figure}
Based on this analysis, it is straightforward to expect that coating of both Co surfaces by graphene should further enhance PMA. Indeed, as shown by the blue stars in Fig.~\ref{fig1}(d), PMA is much strengthened up to twice as much as for the bare case for most Co thicknesses (see corresponding layer resolved contributions represented by blue bars in Fig.~\ref{fig1}(e)-(j)).
In particular, it is interesting to consider the case of 3 ML Co film where PMA remains unaffected when only one surface is coated by graphene [Fig.~\ref{fig1}(d)]. This is because the in-plane anisotropy of the 2$^{nd}$ layer of the Co film is comparably enhanced as well by graphene coating, which compensates the enhanced PMA from the interfacial Co layer [black and red bars in Fig.~\ref{fig1}(e)]. Thus, the coating of graphene on one surface of the 3 ML-thick Co film does not improve the total PMA. However, when both surfaces are coated by graphene, the PMA surprisingly roars up to 2~mJ/m$^2$, which is twice larger than that of bare 3 ML Co film [Fig.~\ref{fig1}(d)].  
The origin of this enhancement is that when both surfaces of the Co film are covered by graphene, the out-of-plane anisotropy of two interfacial Co layers are enhanced and at the same time, the middle layer of Co, which contributes to in-plane anisotropy in bare Co or Co/Gr cases, starts contributing to out-of-plane anisotropy [cf. black, red and blue bars in Fig.~\ref{fig1}(e)]. Therefore, in total, the PMA is almost doubled [Fig.~\ref{fig1}(d)].

In order to elucidate further the mechanism of the PMA enhancement due to the presence of graphene on Co films, we performed a comparison analysis of MAE from orbital-hybridization between surface (interface) Co $3d$ orbitals and bare Co(5ML) or Co(5ML)/Gr films as shown in Fig.~\ref{fig2}.
For bare Co, the hybridization between $d_{xy}$ and $d_{x^2-y^2}$ [labeled as 5 with red bars in Fig.~\ref{fig2}(a)] gives rise to the largest PMA contribution. The second largest part of PMA arises from hybridization between $d_{xz}$ and $d_{yz}$ [labeled as 2 with pink bars in Fig.~\ref{fig2}(a)]. At the same time, the hybridization between $d_{yz}$ and $d_{z^2}$ [labeled as 1 with blue bars in Fig.~\ref{fig2}(a)] constitutes a comparable large in-plane anisotropy. The other hybridizations labeled 3 and 4, [between $d_{xy}$ and $d_{xz}$, and between $d_{yz}$ and $d_{x^2-y^2}$, respectively] give much smaller anisotropy contributions compared to 1, 2 and 5.
When the surface Co atom interacts with graphene, the MAE arising from those hybridizations changes completely [see Fig.~\ref{fig2}(b)]. 
Now, MAE from hybridization 1 contributes to PMA, while the PMA from 5 is strongly reduced compared to the case of pure Co surface. As for hybridization states 3 and 4, the MAE does not change much, which is very similar to the case of Co sandwiched between heavy metals \cite{WangDS}. Thus, the PMA of Co atoms at the interface with graphene is strongly enhanced compared to that at the bare Co surface [see Fig.~\ref{fig1}(f)]. We note that our findings are fully consistent with the recently reported PMA enhancement in C$_{60}$-covered \cite{Repain2015} and graphene-coated Co surfaces \cite{Decker2013}.

As the most important anisotropy change induced by the presence of graphene on Co originates from hybridization between $d_{yz}$ and $d_{z^2}$ orbitals, it is interesting to inspect the graphene's impact on the electronic states corresponding to the Co $3d$ orbitals. In Fig.~\ref{fig2}(c),(d) we plot the projected density of states (pDOS) for surface Co atom in pristine and graphene coated Co films, respectively.
Graphene strongly affects the energy of the different orbitals, one very important change being an inversion of the energy levels of the Bloch states with $3d_{z^2}$ and $3d_{x^2-y^2}$ character due to graphene. Namely, in pristine Co surface $3d_{x^2-y^2}$ states are above $3d_{z^2}$ states (close to the Fermi level), while in case of Co/Gr these states are swapped yielding Co $3d_{z^2}$~states being above $3d_{x^2-y^2}$ [see Fig.~\ref{fig2}(c) and (d), respectively].
This is because the C $2p_z$ states present at the Fermi level in case of pure graphene, strongly hybridize with Co orbitals in case of Co/Gr (mostly the $3d_{z^2}$ states ones for symmetry reasons), causing a shift of $3d_{z^2}$ states towards the Fermi level.
In order to have a direct view of graphene and Co bonding, we also show a cross-section of the charge difference, $\Delta \rho$, (Fig.~\ref{fig2}(e),(f)) calculated by $\Delta \rho$=$\rho$(Co/Gr)-$\rho$(Co)-$\rho$(Gr). We see that the charge accumulation (shown in red) is mainly between Co and graphene interface.
\begin{figure}
  \includegraphics[width=\textwidth]{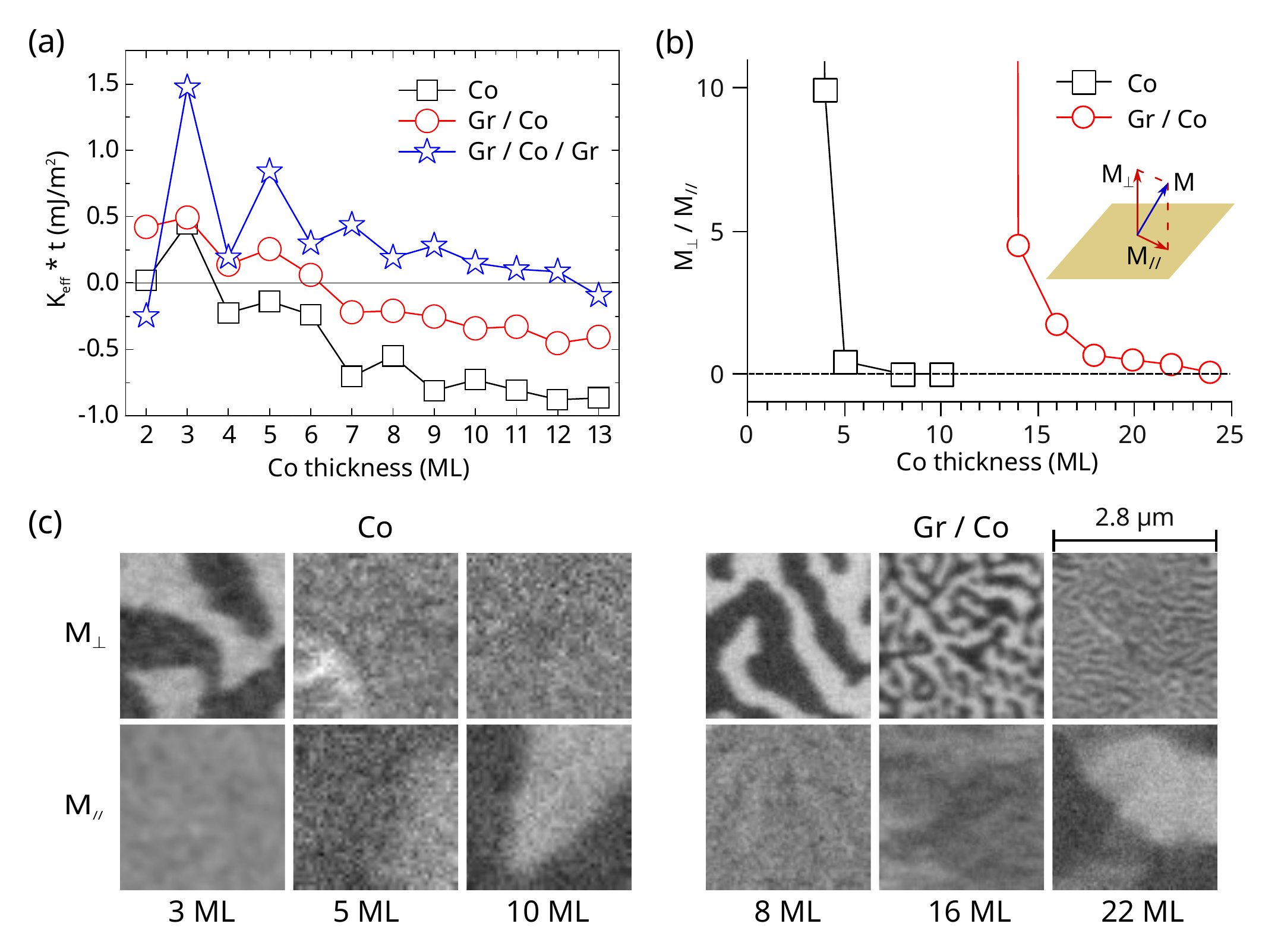}\\
\caption{(a) Effective anisotropy $K_{eff}*t$ as a function of Co thickness for bare Co film (Co), one surface of Co film coated by graphene (Co/Gr) and both surfaces of Co film coated by graphene (Gr/Co/Gr), respectively. (b) Thickness dependence of the $M_\perp$ / $M_{//}$ ratio for bare Co and for graphene / Co films. (c) Out-of-plane ($M_\perp$) and in-plane ($M_{//}$) SPLEEM magnetic images for a 18 ML-thick Co film intercalated between graphene and Ir(111). Field of view is $1.5 \times 1.5 \mu m^2$.}\label{fig3}
\end{figure}

So far, we have discussed the behavior of the surface anisotropy, $K_S$. However, it is very interesting and important to investigate the effective anisotropy $K_{eff}$ in a view of direct comparison to experiments. Here, we calculate $K_{eff}$ assuming the relation $K_{eff}=K_S/t-E_{demag}$, where $t$ is the film thickness and $E_{demag}$ indicates the demagnetization field energy which represents the sum of all the magnetostatic dipole-dipole interactions up to infinity~\cite{Ali}.  
The critical thickness, $t_c$, beyond which $K_{eff}$ changes sign and thus the easy axis switches between perpendicular and parallel to the surface direction, is about 3.7 MLs (7.2\AA) for a bare Co film [Fig.~\ref{fig3}(a)]. 
When coated by graphene on one surface, it is increased up to 6.2 MLs (12.3\AA), which is already comparable to that of Fe/MgO/Fe magnetic tunnel junctions~\cite{Ali}. 
More interestingly, $t_c$ will be further extended up to 12.5 MLs (24.9\AA) in Gr/Co/Gr [Fig.~\ref{fig3}(a) blue stars]. 

Our findings are in good agreement with our experiments.
The magnetic properties of Co thin films directly grown on Ir(111) or intercalated between graphene and an Ir(111) substrate have been explored using spin-polarized low-energy electron microscopy (SPLEEM) \cite{spleem1,spleem2}, a technique available at the National Center for Electron Microscopy of the Lawrence Berkeley National Lab.\cite{Nicolas} All samples were prepared in situ, under ultra-high vacuum conditions (base pressure in the $10^{-11}$ mbar range), in the SPLEEM chamber. An Ir(111) single crystal was used as a substrate and cleaned following a well-established procedure.\cite{sample-prep}. Graphene was subsequently grown by chemical vapor deposition, by exposing the Ir(111) surface to about $5\times10^{-8}$ mbar of C$_2$H$_4$ at 600$^\circ$C. Under these growth conditions, graphene nano-flakes form everywhere on the surface and coat the whole Ir surface in one hour, typically. Cobalt is then evaporated by molecular beam epitaxy at a rate of 0.3 ML per minute while keeping at 300$^\circ$C the previously prepared Gr/Ir(111) interface. At this temperature, Co atoms intercalate below the graphene layer \cite{2012APL, 2012JPCL} without alloying with the Ir(111) surface \cite{Drnec2015}. The in-plane and out-of-plane magnetic domain structures of the intercalated Co film is then revealed, while depositing. An example of SPLEEM images for the Gr/Co/Ir(111) system obtained at room temperature and at remanence is reported in Fig.~\ref{fig3}(b) for a 18 ML-thick Co film. From these two images, the ratio between the component $M_\perp$ of the magnetization perpendicular to the surface and the component $M_{//}$ parallel to the film surface, can be calculated. $M_\perp / M_{//} = 0$ means that magnetization is purely in plane, while $M_\perp / M_{//} \gg 1$ means that magnetization is essentially out-of-plane.

The main result of this experiment is the large difference between the critical thickness at which the out-of-plane to in-plane magnetization reorientation transition occurs for the Co/Ir(111) and graphene/Co/Ir(111) systems. In the former case, magnetization switches from out-of-plane to in-plane at a typical Co thickness of 5 ML. In the latter case however, this change of magnetic anisotropy is observed for much larger Co thicknesses. For a 13 ML thick Co film intercalated between graphene and Ir(111), magnetization is purely out-of-plane (no magnetic contrast is observed when probing the in-plane component of the magnetization. Above 15 ML, the in-plane contrast of the Co film is clearly visible and increases with the Co thickness. At the same time, the out-of-plane contrast continuously decreases, indicating that magnetization rotates towards the film plane as the Co film becomes thicker. Our theoretical finding is thus in a good agreement with our experiments: when in contact with graphene, Co unambiguously favors perpendicular magnetic anisotropy.

\begin{figure}
  \includegraphics[width=\textwidth]{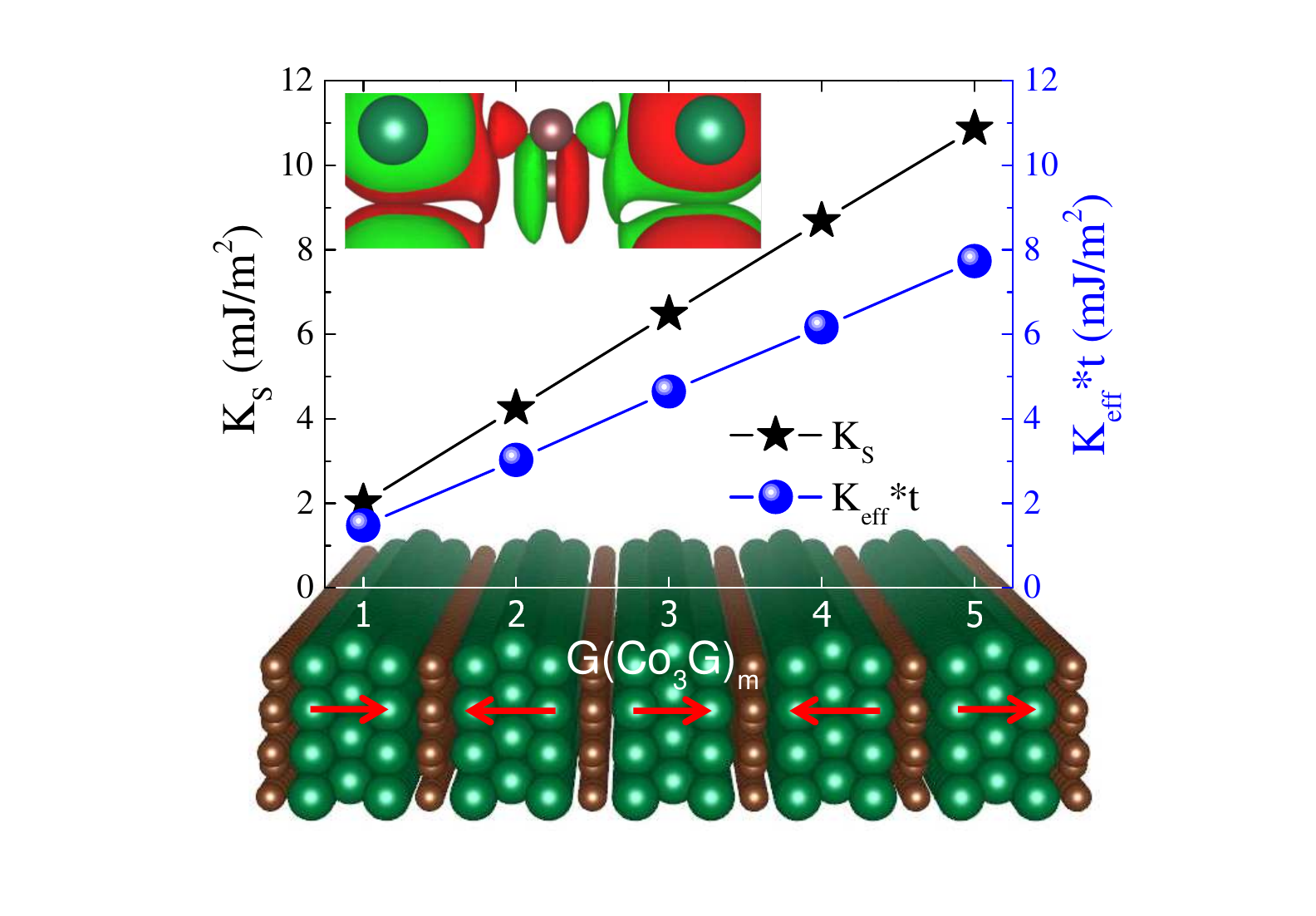}\\
\caption{Surface (interface) magnetic anisotropy, $K_S$ (black stars), and effective anisotropy $K_{eff}*t$ (blue balls) as a function of layers (m) of Co$_3$G super-unit in G(Co$_3$G)$_m$ heterostructure. The red arrows indicate the spin orientation of each the neighboring Co layers next to graphene which are antiferromagnetic coupled to each other via superexchange. Inset is the contour density for spin up (red) and spin down (green) states when the spin density is equal to $\pm $10$^{-3}\mu_B$/\AA$^3$, respectively.}\label{fig4}
\end{figure}

As seen from aforementioned layer-resolved analysis, the PMA of Co-graphene systems is dominated by the first three interfacial Co layers. Taking the advantage of this property, we propose to design Co-graphene heterostructures, G(Co$_n$G)$_m$ comprising $m$ layers of super-unit (Co$_n$G) deposited on a graphene substrate [as shown in Fig.~\ref{fig4} for $n$=3 case].
The ground state of those heterostructures favors anti-parallel coupling between ferromagnetic Co electrodes via graphene layer shown by the contour spin distribution in inset of Fig.~\ref{fig4}, and also indicated by the opposite red arrows in the bottom structure in Fig.~\ref{fig4}. The mechanism responsible for this stability can be attributed to the 180 degree super-exchange coupling between two Co electrodes across graphene. This super-exchange coupling mechanism is consistent with previous study~\cite{superexchange}. One should note, however, that we found that two ferromagnetic Co layers next to graphene favors antiferromagnetic coupling through one carbon atom site while in Ref.~\citenum{superexchange}, the Co electrodes couple to each other through two carbon atoms, 
which is slightly less energetically favorable by about 8 meV compared to that through one carbon site. With these ground states, we found that both the surface anisotropy, $K_S$, and the effective anisotropy, $K_{eff}$*$t$, linearly increase as a function of super-unit number $m$ shown in Fig.~\ref{fig4}. This suggests that Co-graphene heterostructures possess a strong and robust effective PMA. Finally, we would like to point out that the effective PMA value for this heterostructure is more than 20 times stronger than that observed for Co/Pt multilayers~\cite{beach}. In detail, one can see from comparison of $K_{eff}$*$t$ for $m$=3 in Fig.~\ref{fig4} and Fig.~3(c) in Ref.~\citenum{beach}, i.e. 4.6~mJ/m$^2$ for (Co$_3$G)$_3$ versus 0.2~mJ/m$^2$ for (Co$_3$Pt)$_3$.

In conclusion, we demonstrated from both experiments and first-principles calculations that graphene can dramatically enhance the anisotropy of Co films. 
Our calculations showed that the critical thickness switching from out-of-plane to in-plane easy-axis can be extended up to 12~\AA~ and 25~\AA~ depending on coating graphene on one or both surfaces of Co films, respectively. 
Our experimental study of magnetic anisotropy for Co/Ir and graphene/Co/Ir proved that graphene can strongly enhance the PMA of Co surface. 
The mechanism responsible for this anisotropy enhancement is unveiled by our layer- and orbital-hybridization-resolved analysis.
Based on layer-resolved anisotropy analysis, we propose graphene-Co heterostructures and demonstrate that they possess a strong and robust effective PMA which linearly increases as a function of heterostructure thickness. These findings point towards a possible engineering of giant anisotropy graphene-Co heterostructures, which stands as a hallmark for future spintronic information processing technologies.

\acknowledgement

The authors would like to thank B. Dieny, S. Roche and A. Fert for fruitful discussions. The research leading to these results has received funding from the European Union Seventh Framework Programme under grant agreement 604391 GRAPHENE FLAGSHIP, the ANR-2010-BLAN-1019-NMGEM and ANR-12-BS-1000-401-NANOCELLS projects.


\begin{thebibliography}{99}
\bibitem{sample-prep} The Ir(111) single crystal was cleaned in a preparation chamber attached to the SPLEEM using continuous $Ar^+$ ion sputtering, while flashing periodically the surface at $1200^\circ C$ under $10^{-8}$ mbar of oxygen. A last temperature flash ($1200 ^\circ C$) is finally done in the SPLEEM chamber, under UHV conditions, to remove the oxide layer.
\end{thebibliography}
\end{document}